\documentstyle[twoside,fleqn,espcrc2]{article}

\newcommand{\AmS}{{\protect\the\textfont2
  A\kern-.1667em\lower.5ex\hbox{M}\kern-.125emS}}
\def\vep{\varepsilon}
\def\half{{\scriptstyle{1\over 2}}}
\def\thha{{\scriptstyle{3\over 2}}}
\def\tquart{{\scriptstyle{3\over 4}}}
\def\seth{{\scriptstyle{7/3}}}
\newcommand{\real}{\relax{\rm I\kern-.18em R}}
\def\zahlen{{\rm Z \!\! Z}}
\def\Tr{{\rm Tr}}
\def\cO{{\cal O}}
\def\cV{{\cal V}}
\def\cG{{\cal G}}
\def\cA{{\cal A}}
\def\fm{{~\rm fm}}
\def\MeV{{~\rm MeV}}
\def\Ds{{D\hskip-2.2mm\slash}}
\def\min{{\rm min}}

\title{
\vskip-1cm\hfill{\tenrm INLO-PUB-7/97}\vskip1mm  
The QCD vacuum\thanks{Review at Lattice'97, Edinburgh, 22-26 July 1997.}
}
\author{Pierre van Baal
\address{Instituut-Lorentz for Theoretical Physics, University of Leiden,\\
 PO Box 9506, NL-2300 RA Leiden, The Netherlands}}
\begin{document}
\begin{abstract}
We review issues involved in understanding the vacuum, long-distance and 
low-energy structure of non-Abelian gauge theories and QCD. The emphasis will 
be on the role played by instantons.
\end{abstract}
\maketitle
\section{INTRODUCTION}
\vskip-0.7mm 
The term ``QCD vacuum" is frequently abused. Only in the case of the Hamiltonian
formulation is it clear what we mean by the vacuum: it is the wave functional 
associated with the lowest energy state. Observables create excitations on top 
of this vacuum. Knowing the vacuum is knowing all: We should know better. 

Strictly speaking the vacuum is empty. Nevertheless its wave functional can be 
highly non-trivial, deviating considerably from that of a non-interacting Fock 
space, based on a quadratic theory. Even in the later case the result of probing
the vacuum by boundaries is non-trivial as we know from Casimir. The probe is 
essential: one needs to disturb the vacuum to study its properties. Somewhat 
perversely the vacuum may be seen as a relativistic aether. It promises to 
magically resolve our problems, from confinement to the cosmological constant. 
For the latter supersymmetry is often called for to remove the otherwise 
required fine-tuning. It merely hides the relativistic aether, even giving it 
further structure. Remarkably it seems to have enough structure to give a 
non-trivial example of the dual superconductor at work~\cite{Sewi}. 

Most will indeed put their bet on the dual superconductor picture for the QCD 
vacuum~\cite{Dusu}, and this has motivated the hunt for magnetic monopoles 
using lattice techniques, long before supersymmetric duality stole the 
show~\cite{Sewi}. The definitions rely on choosing an abelian 
projection~\cite{Tho1} and the evidence is based on the notion of abelian 
dominance~\cite{Domi}, establishing the dual Meissner effect~\cite{Dume}, or 
the construction of a magnetically charged order parameter, whose non-zero 
expectation value implies spontaneous breaking of the dual gauge 
symmetry~\cite{Digi}, yielding electric confinement. But center vortices, 
probed by the newly defined center dominance, this year suddenly became center 
stage~\cite{Cevo} again and we will surely hear more next year.

In the euclidean path integral only the vacuum state will contribute when we
let the (imaginary) time go to infinity, underlying the essence of the transfer 
matrix approach to extract observables from euclidean path integrals as used 
in lattice gauge theories. However, this is not what people have in mind when 
talking about the vacuum structure of gauge theories. More appropriate for 
most studies would be to talk about the low-energy and long-distance behaviour 
of the theory. One way to address this is to attempt to isolate the relevant 
degrees of freedom for which one can derive an effective low-energy theory. 
Monopole actions derived from block-spin transformations~\cite{Bloc}
and the instanton liquid model~\cite{Shur} are examples.
It is not required that the relevant degrees of freedom are associated to 
semi-classical objects, the main reason being that in a strongly interacting 
theory the quantum fluctuations can be (much) bigger than the classical 
effects. Sometimes this argument is used against the relevance of instantons, 
which by their very definition as localised objects in time might lose 
significance when quantum fluctuations are too strong. 

Due to the limited space available this review concentrates on the instanton 
content of the theory, where recently considerable progress has been achieved. 
But first I will use this opportunity to explain my own thoughts on the matter.

\section{VACUUM DEMOCRACY}

The model we wish to describe here starts from the physics in a small volume,
where asymptotic freedom guarantees that perturbative results are valid.
The assumption is made, that at least for low-energy observables, integrating
out the high-energy degrees of freedom is well-defined perturbatively and
all the non-perturbative dynamics is due to a few low-lying modes. This
is most easily defined in a Hamiltonian setting, since we are interested
in situations where the non-perturbative effects are no longer described
by semiclassical methods. 
\subsection{Complete gauge fixing}
Due to the action of the gauge group on the vector fields, a finite dimensional
slice through the physical configuration space (gauge inequivalent fields) is 
bounded. One way to demonstrate this is by using the complete Coulomb gauge 
fixing, achieved by minimising the $L^2$ norm of the gauge field along the gauge
orbit. At small energies, fields are sufficiently smooth for this to be well 
defined and it can be shown that the space under consideration has a boundary, 
defined by points where the norm is degenerate. These are by definition gauge 
equivalent such that the wave functionals are equal, possibly up to a phase 
factor in case the gauge transformation is homotopically non-trivial. The space
thus obtained is called a fundamental domain. For a review see ref.~\cite{Tren}.

\subsection{Non-perturbative dynamics}
Given a particular compact three dimensional manifold $M$ on which the gauge 
theory is defined, scaling with a factor $L$ allows one to go to larger volumes.
It is most convenient to formulate the Hamiltonian in scale invariant fields 
$\hat A\!=\!LA$. Dividing energies by $L$ recovers the $L$ dependence in the 
classical case, but the need of an ultraviolet cutoff and the resulting scale 
anomaly introduces a running coupling constant $g(L)$, which in the low-energy 
effective theory is the only remnant of the breaking of scale invariance. 

When the volume is very small, the effective coupling is very small and the 
wave functional is highly localised, staying away from the boundaries of the 
fundamental domain. We may compare with quantum mechanics on the circle, seen 
as an interval with identifications at its boundary. At which points we choose 
these boundaries is just a matter of (technical) convenience. The fact that 
the circle has non-trivial homotopy, allows one to introduce a $\theta$ 
parameter (playing the role of a Bloch momentum). 

Expressed in $\hat A$, the shape of the fundamental domain and the nature of 
the boundary conditions, is independent of $L$. Due to the rise of the running 
coupling constant with increasing $L$ the wave functional spreads out over the 
fundamental domain and will start to feel the boundary identifications. This 
is the origin of non-perturbative dynamics in the low-energy sector of the 
theory.

Quite remarkably, in all known examples (for the torus and sphere geometries), 
the sphalerons lie exactly at the boundary of the fundamental domain, with the 
sphaleron mapped into the anti-sphaleron by a homotopically non-trivial gauge 
transformation. The sphaleron is the saddle point at the top of the barrier 
reached along the tunnelling path associated with the largest instanton, its
size limited by the finite volume.

For increasing volumes the wave functional first starts to feel the boundary
identifications at these sphalerons, ``biting its own tail". When the energy 
of the state under consideration becomes of the order of the energy of this 
sphaleron, one can no longer use the semiclassical approximation to describe 
the transition over the barrier and it is only at this moment that the shift
in energy becomes appreciable and causes sizeable deviations from the 
perturbative result. This is in particular true for the groundstate energy. 
Excited states feel these boundary identifications at somewhat smaller volumes,
but nodes in their wave functional near the sphaleron can reduce or postpone 
the influence of boundary identifications. 

This has been observed clearly for $SU(2)$ on a sphere~\cite{Vdhe}. The 
scalar and tensor glueball mass is reduced considerably due to the boundary 
identifications, whereas the oddball remains unaffected (see fig.~1). These 
non-perturbative effects remove an unphysical near-degeneracy in perturbation 
theory (with the pseudoscalar even slightly lower than the scalar glueball 
mass). The dominating configurations involved are associated to instanton 
fields,  in a situation where semiclassical techniques are inappropriate for 
computing the magnitude of the effect. When boundary identifications matter, 
the path integral receives large contributions from configurations that have 
non-zero topological charge, and in whose background the fermions have a 
chiral zero mode, its consequences to be discussed later.

\begin{figure}[htb]
\vspace{2.5cm}
\includegraphics{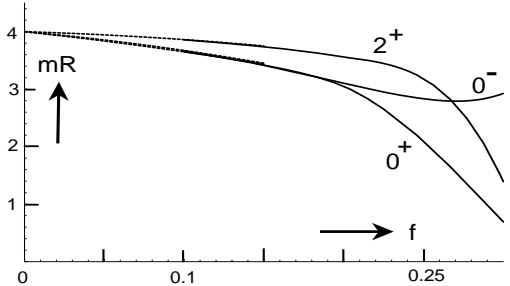}
\caption{The low-lying glueball spectrum on a sphere of radius $R$ as a
function of $f=g^2(R)/2\pi^2$ at $\theta\!=\!0$. Approximately, $f\!=\!0.28$ 
corresponds to a circumference of $1.3\fm$. From ref.~[11].\vskip-7mm}
\label{fig:fig1}
\end{figure}

At some point technical control is lost, since so far only the appropriate 
boundary conditions near the sphalerons can be implemented. As soon as the 
wave functional starts to become appreciable near the rest of the boundary 
too, this is no longer sufficient. 

This method has in particular been very successful to determine the low-lying
spectrum on the torus in intermediate volumes, where for $SU(2)$ agreement with
the lattice Monte Carlo results has been achieved within the 2\% statistical 
errors~\cite{Tren,Kovb}. In this case the non-perturbative sector of the theory
was dominated by the energy of electric flux (torelon mass), which vanishes 
to all orders in perturbation theory. The leading semiclassical result is 
$\exp(-S_0/g(L))$, due to tunnelling through a quantum induced barrier of 
height $E_s\!=\!3.21/L$ and action $S_0\!=\!12.5$. Already beyond $0.1\fm$ 
this approximation breaks down. One finds, accidentally in these small 
volumes, the energy to be nearly linear in $L$. 

The effective Hamiltonian in the zero-momen\-tum gauge fields, derived by 
L\"uscher~\cite{Lusc}, and later augmented by boundary identifications to 
include the non-perturbative effects~\cite{Kovb}, breaks down at the point 
where boundary identifications in the non-zero momentum directions associated 
with instantons become relevant. The sphaleron has an energy $72.605/(Lg^2(L))$
and was constructed numerically~\cite{Gpvb}. Its effect becomes noticeable 
beyond volumes of approximately $(0.75\fm)^3$. For $SU(3)$ this was verified 
directly in a lattice Monte Carlo calculation of the finite volume topological 
susceptibility~\cite{Hoek}. The results for the sphere have shown that also
these effects can in principle be included reliably, but the lack of an 
analytic instanton solution on $T^3\times\real$ has prevented us from doing 
so in practise.

\subsection{Domain formation}
The shape of the fundamental domain depends on the geometry. Assuming that 
$g(L)$ keeps on growing with increasing $L$, causing the wave functional to 
feel more and more of the boundary, one would naturally predict that the 
infinite volume limit depends on the geometry. This is clearly unacceptable, 
but can be avoided if the ground state obtained by adiabatically increasing 
$L$ is not stable. Thus we conjecture that the vacuum is unstable against 
domain formation. This is the minimal scenario to make sure that at large 
volumes, the spectrum is independent of its geometry. Domains would naturally 
explain why a non-perturbative physical length scale is generated in QCD, 
beyond which the coupling constant will stop running. However, we have no 
guess for the order parameter, let alone an effective theory describing
excitations at distances beyond these domains. Postulating their existence,
nevertheless a number of interesting conclusions can be drawn.

The best geometry to study domain formation is that of a box since it is 
space-filling. We can exactly fill a larger box by smaller ones. This is not 
true for most other geometries. In small to intermediate volumes the vacuum 
energy density is a decreasing function~\cite{Kovb} of $L$, but in analogy to 
the double well problem one may expect that at stronger coupling the vacuum 
energy density rises again with a minimum at some value $L_0$, assumed
to be $0.75\fm$.  For $L$ sufficiently larger than $L_0$ it thus becomes 
energetically favourable to split the volume in domains of size $L_0^3$. 

Since the ratio of the string tension to the scalar glueball mass squared shows
no structure around $(0.75\fm)^3$, we may assume that both have reached their 
large volume value within a domain. The nature of their finite size corrections
is sufficiently different to expect these not to cancel accidentally. The 
colour electric string arises from the fact that flux that enters the box has 
to leave it in the opposite direction. Flux conservation with these building 
blocks automatically leads to a string picture, with a string tension as 
computed within a single domain and a transverse size of the string equal to 
the average size of a domain, $0.75\fm$. The tensor glueball in an intermediate
volume is heavily split between the doublet ($E^+$) and triplet ($T^+_2$)
representations of the cubic group, with resp. 0.9 and 1.7 times the scalar 
glueball mass. This implies that the tensor glueball is at least as large as 
the average size of a domain. Rotational invariance in a domain-like vacuum 
comes about by averaging over all orientations of the domains. This is expected
to lead to a mass which is the multiplicity weighted average of the doublet 
and triplet, yielding a mass of 1.4 times the scalar glueball mass. Domain 
formation in this picture is {\em driven} by the large field dynamics 
associated with sphalerons. Which state gets affected most depends in an 
intricate way on the behaviour of the wave functionals (cmp. fig.~1).

In the four dimensional euclidean context, $O(4)$ invariance makes us assume
that domain formation extends in all four directions. As is implied by 
averaging over orientations, domains will not neatly stack. There will be 
dislocations which most naturally are gauge dislocations. A point-like gauge 
dislocation in four dimensions is an instanton, lines give rise to monopoles
and surfaces to vortices. In the latter two cases most naturally of the  $Z_N$ 
type. We estimate the density of these objects to be one per average domain 
size. We thus predict an instanton density of $3.2\fm^{-4}$, with an average 
size of $1/3\fm$. For monopoles we predict a density of $2.4\fm^{-3}$. 

If an effective colour scalar field will play the role of a Higgs field, abelian
projected monopoles will appear. It can be shown~\cite{Taub} that a monopole (or
rather dyon) loop, with its $U(1)$ phase rotating $Q$ times along the loop
(generating an electric field), gives rise to a topological charge $Q$. 
In abelian projection it has been found that an instanton always contains a 
dyon loop~\cite{Inmo}. We thus argue this result to be more general, leading
to further ties between monopoles and instantons.

\subsection{Regularisation and $\theta$}
It is useful to point out that the non-trivial homotopy of the physical 
configuration space, like non-contractable loops associated to the instantons 
($\pi_1(\cA/\cG)\!=\!\pi_3(G)\!=\!\zahlen$), is typically destroyed by the 
regularisation of the theory. This is best illustrated by the example of 
quantum mechanics on the circle. Suppose we replace it by an annulus. As long 
as the annulus does not fill the hole, or we force the wave function to vanish 
in the middle, theta is a well-defined parameter associated to a multivalued 
wave function. We may imagine the behaviour for small instantons in gauge 
theories to be similar to that at the center in the above model. Indeed, the 
gauge invariant geodesic length of the tunnelling path for instantons on 
$M\times\real$, given by $\ell=\int^\infty_{-\infty}dt\sqrt{\cV(t)}$, where 
$\cV(t)$ is the classical potential along the tunnelling path, is expected to 
vanish for instantons that shrink to zero size. This is confirmed for 
$M\!=\!S^3$, using results of ref.~\cite{Dass} to show that (for unit radius) 
$\ell\!=\!2\Gamma^2(\tquart)\sqrt{6\pi/b}(1\!+\!\cO(b^{-\thha}))$, with 
the instanton size defined by $\rho\!\equiv\!(1\!+\!b^2)^{-\half}$.

Due to the need of regularising the ultraviolet behaviour, the small instantons
are cut out of the theory. Using the lattice regularisation this does not leave
a hole, but rather removes the ``singularity at the origin", as the lattice
configuration space has no non-contractable loops. Strictly speaking this means
one can not have a theta parameter at any finite lattice spacing. Furthermore 
the regularisation forces one to divide out all the gauge transformations as 
there are no homotopically non-trivial ones. It is advisable to divide 
out {\em all} gauge transformations, even if some of the homotopy is preserved 
by some regularisation! 

One can, however, still introduce the theta parameter by adding $i\theta Q$ 
(with $Q\!=\!\int d_4x q(x)$, $q(x)\!=\!Tr(F^{\mu\nu}(x)\tilde F_{\mu\nu}(x))
/16\pi^2$ the topological charge operator) to the action. Of course only for 
smooth fields the charge $Q$ will be (approximately) integer. Also within the 
Hamiltonian formulation one may introduce a theta parameter by mixing the 
electric field with $\theta$ times the magnetic field, $E\rightarrow E-\theta 
B/(2\pi)^2$. In these approaches theta is simply a parameter added to the 
theory. Whether or not one will retrieve the expected periodic behaviour in 
the continuum limit becomes a dynamical question. 

It should be pointed out that in particular for $SU(N)$ gauge theories in a box 
(in sectors with non-trivial magnetic flux) there is room to argue for a 
$2\pi N$, as opposed to a $2\pi$, periodicity for the $\theta$ dependence.
However, the spectrum is periodic with a period $2\pi$, and the apparent 
discrepancy is resolved by observing that there is a non-trivial spectral 
flow~\cite{Tho2}. This may lead to phase transitions at some value(s) of 
$\theta$, related to the oblique confinement mechanism~\cite{Tho1}. Similarly 
for supersymmetric gauge theories this interpretation, supported by the recent 
discovery of domain walls between different vacua~\cite{Dowa}, removes the 
need for semiclassical objects with a charge $1/N$. Such solutions do exist 
for the torus, but the fractional charge is related to magnetic flux and the 
interpretation is necessarily as stated above! The ``wrong" periodicity in 
theta has long been used to argue against the relevance of instantons, but in 
the more recent literature this is now phrased more cautiously~\cite{Shif,Zhit}.

\section{INSTANTONS}
Instantons are euclidean solutions responsible for the axial anomaly, breaking 
the $U_A(1)$ subgroup of the $U(N_f)\times U(N_f)$ chiral symmetry for $N_f$ 
flavours of massless fermions~\cite{Tho3}, as dictated by the anomaly,
$\sum_f\partial_\mu\bar\Psi_f\gamma^\mu\gamma_5\Psi_f(x)\!=\!2N_f q(x)$. The 
breaking of $U_A(1)$ manifests itself in the semiclassical computations through
the presence of fermion zero modes, with their number and chirality fixed by 
the topological charge, through the Atiyah-Singer index theorem~\cite{Atiy}.
Integration over the fermion zero modes leads to the so-called 't Hooft
vertex or effective interaction~\cite{Tho3}. The integration over the scale 
parameter of the instanton ensemble is infrared dominated and a 
non-perturbative computation is desirable. 

In addition it is believed that the instantons are responsible for chiral 
symmetry breaking, where a chiral condensate is formed, which breaks the axial 
gauge group $U_A(N_f)$ completely. This spontaneous breaking is dynamical and 
it is less well established that instantons are fully responsible. It is the 
basis of the instanton liquid model as developed by Shuryak over the years. 
For a comprehensive recent review see ref.~\cite{Shur}. The details of the 
instanton ensemble play an important role. Only a liquid-like phase, as opposed
to the dilute or crystalline phases, will give rise to a chiral condensate. The 
model also makes a prediction for the average size and the topological 
susceptibility. In particular the latter quantity should be well-defined 
beyond a semiclassical approximation. For large sizes the instanton 
distribution is exponentially cut-off and instantons do not give rise to an 
area law for the Wilson loop. When large instantons are more weakly suppressed 
the situation may differ~\cite{Fuku}, but a semiclassical analysis in this 
case should not be trusted.
\vskip-0.2mm
Remarkably the topological susceptibility in pure gauge theories can be related
to the $\eta^\prime$ mass through the so-called Witten-Veneziano relation, 
$f_\pi^2(m_\eta^2\!+\!m_{\eta^\prime}^2\!-\!2m_K^2)/2N_f\!=\!\int\!d_4x\!
<\!T(q(x)q(0))\!>_R$ $\equiv\!\chi_t$, leading to the prediction 
$\chi_t\!\sim\!(180\MeV)^4$. This is based on the fact that the $U_A(1)$ 
symmetry is restored in the planar limit~\cite{Witt,Vene}, with $\chi_t$ of 
order $1/N^2$. From the requirement that in the presence of massless quarks 
$\chi_t$ (and the theta dependence) disappears, the pure gauge susceptibility 
can be related to the quark-loop contributions in the pseudoscalar channel. 
Pole dominance requires the lightest pseudoscalar meson to have a mass squared 
of order $1/N$. Relating the residue to the pion decay constant gives the 
desired result~\cite{Vene}. The index $R$ indicates the necessity of 
equal-time regularisation~\cite{Witt}. A derivation on the lattice using 
Wilson and staggered fermions was obtained in ref.~\cite{Smvi}, making use of 
Ward-Takahashi identities. Finally, also the coarse grained partition 
function of the instanton liquid model~\cite{Shur} allows one to directly 
determine the 't Hooft effective Lagrangian~\cite{Tho3}, from which the 
Witten-Veneziano formula can be read off~\cite{Verb}. This formula is almost 
treated as the holy grail of instanton physics. It is important to realise 
that some approximations are involved, although it is gratifying there are 
three independent ways to obtain it~\cite{Witt,Vene,Smvi,Verb}.
\subsection{Field theoretic method}
A direct computation of $\chi_t\!=\!\int\!\!d_4x<\!q(x)q(0)\!>$ on the lattice 
requires a choice of discretisation for the charge density. A particularly 
simple one is~\cite{Dive} $q_L(x)\!=\!-\!\sum Tr(U_{\mu\nu}(x)\tilde U_{\mu\nu}
(x))/16\pi^2$, where $U_{\mu\nu}(x)\!=\!\half\vep_{\mu\nu\rho\sigma}\tilde 
U_{\rho \sigma}(x)$ is the clover averaged plaquette $P_{\mu\nu}(x)$, formed by
the four plaquettes that meet at $x$. Due to the short distance singularities 
the operators require renormalisations. The lattice charge operator is 
conjectured to satisfy $Q_L\!=\!\sum_x q_L(x)\!=\!a^4QZ(\beta)\!+\!\cO(a^6)$, 
correcting for the fact that $Q_L$ need not be integer. In addition an additive
renormalisation, associated to the contact term discussed before~\cite{Witt}, 
arises due to mixing of $\chi_L$ with operators with the same quantum numbers, 
$\chi_L\!=\!a^4\chi_tZ^2(\beta)\!+\!M(\beta)\!+\!\cO(a^6)$.

To determine $Z(\beta)$ and $M(\beta)$ one takes a classical configuration with
a fixed topological charge $Q$~\cite{Alle}. Monte Carlo updates are rejected if
they change the charge as determined by cooling and subsequent measurements of 
$Q_L$ and $\chi_L$ allows one to extract $Z$ and $M$. (In a sector with fixed
charge $Q$ we note that $\chi_t\!=\!Q^2/V$, with $V$ the volume.) Over a certain
range, independence of the starting configuration and the volume has been 
observed and was henceforth assumed~\cite{Alle}. Due to the need to fix $Q$ to
determine the renormalisation factors, this is in a sense a hybrid method, and 
might also be sensitive to some of the problems of cooling - to be discussed 
shortly. 

Considerable progress has been achieved, however, by repeating the study with 
other choices for $q_L$, based on smearing the links (iteratively replacing 
them by staples). This considerably reduces the value of $M$, greatly 
facilitating the extraction of a signal. On a $16^4$ lattice one finds 
$\chi_t\!=\!(175(5)\MeV)^4$ (for $SU(3)$ at $\beta\!=\!5.9$, 6.0 and 6.1) and 
$\chi_t\!=\!(198(8)\MeV)^4$ (for $SU(2)$ at $\beta\!=\!2.44$, 2.5115 and 2.57).
For the discussion on finite temperature, how the scale was set and for 
further details and references see ref.~\cite{Alle}.
\subsection{Cooling}
In the continuum, the Schwartz inequality implies that the action of any field 
configuration with charge $Q$ is bounded by $8\pi^2|Q|$, and is reached by 
(anti-)selfdual solutions. In its simplest form, cooling aims at finding this 
minimal action, using it to identify the topological charge. As all quantum 
fluctuations are removed, no renormalisations are required, such that 
$\chi_t\!=<\!Q^2\!>\!/V$, where the average is over the Monte Carlo 
ensemble. Cooling can be achieved by putting $\beta=\infty$ in the standard 
Monte Carlo update (accepting updates only if they lower the 
action). The same result is obtained by a sort of congruent 
gradient method, which uses the lattice equations of motion to lower the 
action~\cite{Bete}. For $SU(2)$ this method is deterministic and allows for 
estimating its rate of convergence~\cite{Gpvb}.

It is a remarkable and deep mathematical property of non-Abelian gauge theories
on a compact four dimensional manifold that exact $SU(N)$ solutions exist for 
any charge $Q$ with $4N|Q|$ parameters (moduli), in general describing position,
size and colour orientation of $|Q|$ pseudoparticles. Although a compact 
manifold breaks the scale invariance, generically instantons with arbitrary 
size exist, only limited by the finite volume. A notable exception is charge 
one instantons on $T^4$. There is no problem in having smooth configurations 
of unit charge, but in an attempt to make them self-dual they shrink to 
a point~\cite{Braa}.

To understand its implications we take the time direction to infinity, in which
case finite action forces tunnelling from vacuum to vacuum configuration. The 
vacuum on a torus is, however, not unique. Periodic boundary conditions in the 
time direction force the vacua at both ends to be the same and as soon as one 
releases this constraint, exact vacuum to vacuum tunnelling solutions exist. 
This was studied for $O(3)$ through the exact solutions on a 
cylinder~\cite{Snip}. For gauge theories it can be proven that twisted (for 
$SU(2)$ ``anti-periodic") boundary conditions~\cite{Tho2} remove the 
obstruction, even at finite $T$. Large instantons have finite size 
effects~\cite{Oimc} of $\cO(1/T)$. For sufficiently large volumes, assuming 
the instanton size will be cutoff dynamically, the effect is irrelevant, 
further helped by the fact that in a large volume almost all configurations 
have higher charges.

As always, the continuum limit competes with the infinite volume limit. Even 
with presentday computer power, the remaining window is uncomfortably small.
The above finite size effect is usually swamped by the cutoff effects. As the 
lattice cutoff breaks the scale and rotational invariance, we would expect 
that the action is no longer constant on the continuum moduli space. Indeed,
for a smooth instanton the Wilson action behaves as $S_W(\hat\rho\!\equiv
\!\rho/a)\!=\!8\pi^2(1\!-\!\hat\rho^{-2}/5 \!+\!\cO(\hat\rho^{-4}))$ and 
causes the instanton to shrink, until it becomes of the size of the cutoff 
and falls through the lattice. Cooling will first remove high-frequency modes 
and one is left with a slow motion along the moduli space, giving rise to a 
plateau in the cooling history, used to identify the topological charge. One 
will miss instantons smaller than some fixed value $\hat\rho_c$. Assuming 
asymptotic freedom, one easily shows that the error vanishes in the continuum 
limit. Note that by construction, the cooling method never will associate 
charge to a dislocation with an action smaller than $96\pi^2/11N$, the 
entropic bound, which would spoil scaling~\cite{Disl}.

For extracting the size distribution, cooling and under-relaxed (or slow) 
cooling~\cite{Mich} is problematic as the size clearly will depend on where 
along the plateau one analyses the data~\cite{Brow}. The size distribution can 
be made to scale properly only at the expense of carefully adjusts the number 
of cooling steps~\cite{Smit} when going to different $\beta$.

One can avoid loosing instantons under cooling by modifying the action such 
that the scaling violations change sign~\cite{Oimc}, for example by adding 
a $2\!\times\!2$ plaquette to the Wilson action. This so-called over-improved 
action has the property that instantons grow under cooling, until stopped by 
the finite volume. Consequently it would still mutilate the size distribution.
This can be avoided by improving the action so as to minimise the scaling 
violations~\cite{Mgpe}. A particularly efficient choice is the so-called 
five-loop improved ($5Li$) lattice action: 
$S_{5Li}\!=\!\sum_{m,n}c_{m,n}\sum_{x,\mu\nu}\Tr(1\!-\!P_{m\mu,n\nu}(x))$, 
where $P_{m\mu,n\nu}(x)$ is the $m\!\times\!n$ plaquette and 
$c_{1,1}\!=\!{\scriptstyle{65\over 36}}$, 
$c_{2,2}\!=\!-{\scriptstyle{11\over 720}}$, 
$c_{1,2}\!=\!-{\scriptstyle{8\over 45}}$, 
$c_{1,3}\!=\!{\scriptstyle{1\over 90}}$ and
$c_{3,3}\!=\!{\scriptstyle{1\over 1620}}$. 
Overimproved cooling~\cite{Oimc} is defined by $c_{1,1}\!=\!(4\!-\!\vep)/3$ and
$c_{2,2}\!=\!(\vep\!-\!1)/48$, which is $\cO(a^2)$ improved for $\vep\!=\!0$.
The $5Li$ is improved to $\cO(a^4)$ and allows for a tiny action barrier of 
less than one permill of the instanton action at $\hat\rho_c=2.3$. Instantons 
of smaller size will be rapidly lost under cooling, but larger ones will stay 
forever and practically not change their size, of course using twisted boundary
conditions. For $\hat\rho\!>\!4$ the action is at most a factor $10^{-4}$ 
larger than its continuum value.  After thus eliminating the cutoff effects, 
the finite size effects in the charge one sector are clearly observed, see 
fig.~2.

\begin{figure}[htb]
\vspace{2cm}
\includegraphics{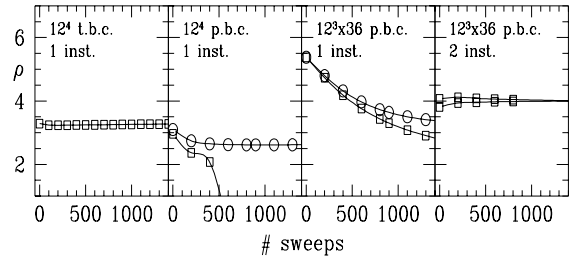}
\caption{Cooling history~[40] for $SU(2)$ charge one (and two) instanton with 
twisted and periodic boundary conditions. Squares represent $5Li$ and 
circles over-improved cooling ($\vep\!=\!-1$).\vskip-8mm}
\label{fig:fig2}
\end{figure}

It is important to note that improved cooling~\cite{Mgpe} is used as a 
diagnostic tool; the configurations are still generated by the standard Wilson
action. One may of course also use improved actions for this purpose, but the 
$5Li$ action was simply not tuned dynamically. One difficulty in extracting 
the instanton distribution is that a typical ensemble will have both instantons
and anti-instantons. These interact, although the action only slightly differs
from $8\pi^2$ times the number of pseudoparticles, provided they are not too
close. Close instanton anti-instanton (I-A) pairs will annihilate. Cooling 
sufficiently long also removes more widely separated pairs. As the cooling makes
the configuration smooth, the total charge can be measured reliably even in the 
presence of these pairs. For SU(2), lattices of sizes $12^4$ and $12^3\!\times
\!36$ at $\beta\!=\!2.4$ and 2.5 with periodic and twisted boundary conditions,
as well as $24^4$ at $\beta\!=\!2.6$ with twisted boundary conditions were used.
At $\beta\!=\!2.5$ a physical volume of $1.02\fm$ across gives sizable finite 
size corrections. One obtains $\chi_t\!=\!(200(15)\!\MeV)^4$ with good scaling
properties~\cite{Mgpe}, which agrees with ref.~\cite{Alle}.

Extracting the size of an instanton is based on identifying the pseudoparticles.
Using among other things the known profile of an isolated instanton, five 
different definitions for the size $\rho$ were used, which are all well 
correlated~\cite{Mgpe}. The resulting size distribution is given in fig.~3,
combining results for $\beta\!=\!2.4$ (averaged over the different lattice 
types and boundary conditions after 20 cooling sweeps) and for $\beta\!=\!2.6$ 
(after 50 cooling sweeps). The solid curve is a fit to the formula $P(\rho)
\propto\rho^{\seth}\exp(-(\rho/w)^p)$, with $w\!=\!0.47(9)\fm$ and $p\!=\!3(1)$, which at small sizes coincides with the semiclassical result~\cite{Tho3}. 
The peak of this distribution occurs at $\rho\!=\!0.43(5)\fm$. Under prolonged 
cooling, up to 300 sweeps, I-A instanton annihilations and in particular finite
size effects in the charge one sector do affect the distribution somewhat, but 
not the average size, which therefore seems to be quite a robust result. 

\begin{figure}[htb]
\vspace{3.8cm}
\includegraphics{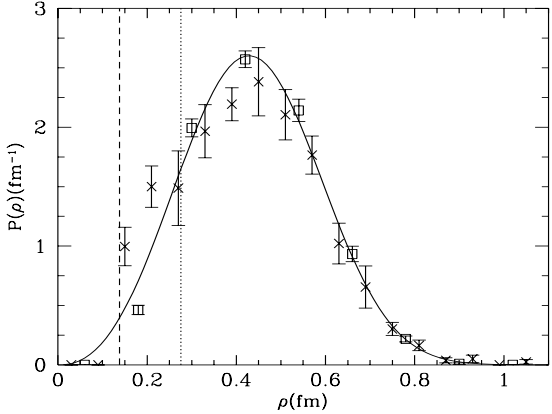}
\caption{$SU(2)$ instanton size distribution for $\beta\!=\!2.4$ (squares) and
2.6 (crosses) in a volume $1.44\fm$ across at lattice spacings $a\!=\!0.12$ 
and $0.06\fm$. The dotted and dashed lines represent the cutoff at 
$a\hat\rho_c$ for both lattices. From ref.~[40].\vskip-7mm}
\label{fig:fig3}
\end{figure}

It would be advantageous if one could come up with a definition for the size 
that is related to a physical quantity, since now the notion is based on the
semiclassical picture. This is nevertheless appropriate for the comparison
with the instanton liquid. The relatively large value of the average size 
as compared to that of $1/3\fm$ predicted by the instanton liquid~\cite{Shur}
is a point of worry, typically leading to stronger interactions that may
lead to a crystal (without chiral symmetry breaking), rather than a liquid.
Nevertheless, in ref.~\cite{Mgpe} it has been tested that the pseudoparticles
are homogeneously distributed with a density of $2\!-\!3\fm^{-4}$ and occupying 
nearly half the volume. This is the case when only close I-A pairs have
annihilated and therefore depends on the amount of improved cooling. It does,
however, show that the pseudoparticles are relatively dense (more so than
assumed in the instanton liquid~\cite{Shur}). The value of $3.2\fm^{-4}$
for the density, derived earlier in the context of the domain picture is
quite realistic in the light of these results.
\subsection{Smoothing}
Another method to study instantons on the lattice is based on the classical 
fixed point actions~\cite{Hani}, defined through the saddle point equation 
$S^{FP}(V)\!=\!\min_{\{U\}}(S^{FP}(U)\!+\!\kappa T(U,V))$.
It is obtained as the weak coupling limit of the blocking transformation
with a positive definite kernel $\kappa T(U,V)$, which maps a lattice $\{U\}$
to $\{V\}$, coarser by a factor two. Reconstructing the fine from the coarse 
lattice is called inverse blocking. 

It can be shown to map a solution of the lattice equations of motion to a 
solution on the finer lattice with the same action. Iterating the inverse
blocking, the lattice can be made arbitrarily fine, thereby proving the
absence of scaling violations to any power in the lattice spacing~\cite{Hani}. 
This classically perfect action still looses instantons below a critical
size, which is typically smaller than a lattice spacing. For solutions this 
most likely happens at the point where the continuum interpolation of the 
lattice field is ambiguous, causing the integer geometric charge~\cite{Geom} 
to jump. For rough configurations that are not solutions, inverse blocking 
typically reduces the action by a factor 32 and makes it more smooth. The 
fixed point topological charge is defined as the limiting charge after 
repeated inverse blockings. This guarantees no charge will be associated 
to dislocations (of any action below the instanton action).

The classical fixed point action, although optimised to be short range, still
has an infinite number of terms and finding a suitable truncation is a practical
problem. From examples of parametrisations, the success of reducing scaling 
violations in quantities like the heavy quark potential (tested by restoring 
rotational invariance) is evident, for recent reviews see ref.~\cite{Impr}. 
In practise only a limited number of inverse blockings is feasible and the 
fixed point topological charge has to rely on a rapid convergence. The closer 
one is able to construct the fixed point action the better this convergence 
is expected to be. For two dimensional non-linear sigma models sufficient 
control was achieved to demonstrate that more than one inverse blocking did 
not appreciably change the topological charge~\cite{Twod}.

In four dimensional gauge theories, both finding a manageable parametrisation 
and doing repeated inverse blockings is a major effort. It goes without saying 
that if no good approximation to the fixed point action is used, one cannot rely
on its powerful theoretical properties. Studies of instantons for $SU(2)$ gauge 
theories were performed in ref.~\cite{Dghz}. A 48 term approximation to the 
fixed point action was used to verify the theoretical properties. The geometric
charge was measured after one inverse blocking and it was shown that for 
$Q\!=\!1$ the instanton action was to within a few percent from the continuum 
action (slightly above it due to finite size effects), whereas for $Q\!=\!0$ 
the action was always lower. Subsequently a simplified eight parameter form was
used on which the instanton action was somewhat poorly reproduced, but such 
that the $Q\!=\!1$ boundary stayed above the entropic bound for the action.

A value of $\chi_t\!=\!(235(10)\MeV)^4$ was quoted on a $8^4$ lattice with 
physical volumes of up to $1.6\fm$, taking full advantage of the fact that
fixed point actions can be simulated at rather large lattice spacings. However,
$<\!Q^2\!>$ measured on the coarse lattice was up to a factor 4 larger than on 
the fine lattice (for the two dimensional study much closer agreement was 
seen~\cite{Twod}). Further inverse blocking to check stability of the charge 
measurement was not performed. 

The same eight parameter action was used in ref.~\cite{Berl}, but they did 
their simulations on the fine lattice and performed an operation called 
smoothing: first blocking and then inverse blocking. They changed the 
proportionality factor $\kappa$ in the blocking kernel, requiring that the 
saddle point condition is satisfied for the blocked lattice. Due to the change
of $\kappa$ the properties of the fixed point action that inspired these 
authors can unfortunately no longer be called upon as a justification.
This smoothing satisfies the properties of cooling (the action always decreases 
and stays fixed for a solution) and should probably by judged as such. (See for 
further comments below.) They consider their study exploratory and concentrate 
on finite temperature near and beyond the deconfinement transition.

In ref.~\cite{Dghk} the number of terms to parametrise the fixed point action 
was extended to four powers of resp. the plaquette, a six-link and an eight-link
Wilson loop. The latter was required to improve on the properties for the 
classical solutions. They achieved $\hat\rho_c\!=\!0.94$, still considerably 
smaller than for $S_{5Li}$, and reproduced the continuum instanton action to 
a few percent for $\rho\!>\!\rho_c$. To increase the quality of the fit a 
constant was added, which should vanish in the continuum limit (as it drops 
out of the saddle point equation). Possible ramifications of this at finite 
coupling are not yet sufficiently understood. 

After one inverse blocking insufficient smoothing is achieved to extract the 
pseudoparticle positions and sizes and further inverse blocking was considered 
computationally too expensive. Like in ref.~\cite{Berl}, they also introduced a 
smoothing cycle, but now by blocking the fine lattice back to the coarse one. 
Such a cycle would {\em not} change the action when the blocking is indeed to 
the {\em same} coarse lattice. However, there are $2^4$ different coarse 
sublattices associated to a fine one and in ref.~\cite{Dghk} the smoothing 
cycle involved blocking to the coarse lattice shifted along the diagonal over 
one lattice spacing on the fine lattice. Unlike in ref.~\cite{Berl}, the 
smoothing cycle will be repeated.

\begin{figure}[htb]
\vspace{1.8cm}
\includegraphics{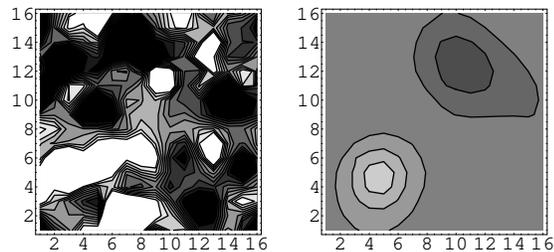}
\caption{Example of the smoothing after 1 and 9 cycles, shown on the fine
lattice. From ref.~[47].\vskip-8mm}
\label{fig:fig4}
\end{figure}

Although the fixed point nature
of the action guarantees it is close to a perfect classical action it needs to 
be demonstrated that it preserves the topological charge at sufficiently large 
scales. For cooling this is argued from the local nature of the updates, not 
affecting the long distance behaviour. Improved cooling is in this sense less 
local than ordinary cooling, and it might seem that due to the rather compact 
Wilson loops involved in the parametrisation of the fixed point action, the 
situation for the smoothing cycle is intermediate. Nevertheless, it should be 
pointed out that the global minimisation involved in inverse blocking, at least
naively, has its effect felt over the entire lattice. 

As evidence in favour of preserving long range physics, it was shown~\cite{Dghk}
that the string tension is conserved under this smoothing (unlike for the method
of ref.~\cite{Berl}, that reports changes up to 25\%). That smoothing is 
successful in removing noise is seen in fig.~4. Somewhat surprisingly close I-A
pairs remained stable for distances as close as 80\% of the sum of their radii. 
Virtually no change is seen for up to 9 smoothing cycles. As I-A pairs are not 
solutions one would have expected some change. Possibly this is due to 
critical slowing down as can also occur for cooling~\cite{Gpvb}. It depends
quite intricately on the details of the mapping implied by the smoothing cycle.

\begin{figure}[htb]
\vspace{3.8cm}
\includegraphics{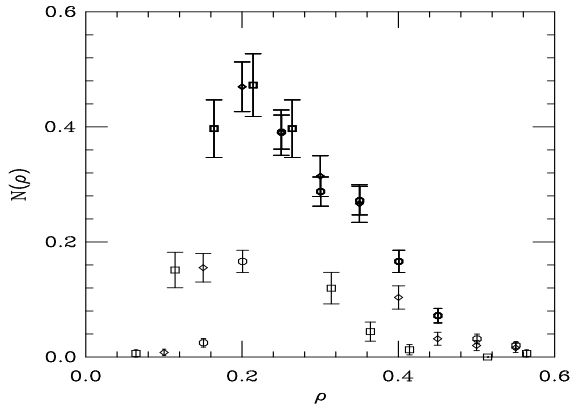}
\caption{The $SU(2)$ integrated instanton size distribution (integrated 
over bins of size $0.05\fm$) for $\beta\!=\!1.4$ (octagons), 1.5 (diamonds) 
and 1.6 (squares) on a $8^4$ lattice, with resp. $a\!=\!0.188$, $0.144$ and 
$a\!=\!0.116\fm$. From ref.~[47].\vskip-6mm}
\label{fig:fig5}
\end{figure}

In fig.~5 the resulting instanton size distribution is given and we refer to
ref.~\cite{Dghk} for the details on the analysis. The instanton density is 
approximately $2.0\fm^{-4}$ and the size peaks at $0.2\fm$. This points to a 
rather dilute situation, although a study of correlations among the instantons 
seems to point to clustering. The susceptibility $\chi_t\!=\!(230(10)\MeV)^4$
agrees with the earlier value~\cite{Dghz}. Cutting out instantons below 
$0.27\fm$ (see fig.~3) would give $\chi_t\!=\!(190\MeV)^4$, but a cut at half 
this value would cause no significant change. It would imply large scaling
violations for the improved cooling method, which were not observed in 
ref.~\cite{Mgpe}. Both the susceptibility and the average size therefore 
disagree significantly with improved cooling.

\subsection{Spectral flow}
One can extract the topological charge by counting the number of chiral zero 
modes of the Dirac operator, using the Atiyah-Singer index theorem~\cite{Atiy}.
Although only valid for smooth configurations, a lattice version could in 
principle be defined~\cite{Vink}, $Q\!=\!m\kappa_P\Tr(\gamma_5/(\Ds\!+\!m))/
N_f$, with $\kappa_P$ a renormalisation constant that depends on the 
lattice definition of the Dirac operator $\Ds$ and $\gamma_5$ used. Due to the 
discretisation of the lattice no exact zero modes exist. For Wilson fermions 
these would-be zero modes will typically be real and give rise to poles in the 
euclidean path integral that are related to the ``exceptional configurations''. 
Given the relation to instantons care is required in handling these 
configurations and in a recent series of papers a ``pole shifting'' algorithm 
has been proposed to remedy this problem~\cite{Bard}. One might also be 
tempted to extract the index by counting the real eigenvalues~\cite{Lang}.

Inspired by the overlap formulation for chiral fermions~\cite{Nane}
a much simpler method to extract the zero modes has been recently 
used~\cite{Navr}. The Wilson Dirac operator has the property that 
$H(U)\!\equiv\!\gamma_5(\Ds(U)\!-W(U)+4-\!m)$ is hermitian. In the continuum 
this operator has a spectral flow as a function of $m$. Non-zero modes cross
zero in pairs of opposite chirality (one going up, the other going down). A 
zero mode is generically responsible for an isolated crossing (the direction of
crossing determines the chirality). It is these properties that are expected to
be robust under discretisation~\cite{Nane}. All that should change on the 
lattice is that the zero modes no longer cross at $m\!=\!0$ (we know that $m$ 
needs to be tuned to $m_c\!\neq\!0$ at finite $\beta$, but for individual 
configurations the crossings will occur at different values of $m$). The 
topological charge is now simply defined as the number of net crossings.

\begin{figure}[htb]
\vspace{3.5cm}
\includegraphics{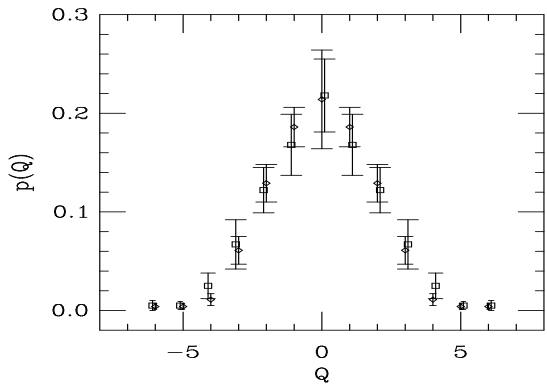}
\caption{Comparison of the $SU(2)$ topological charge distribution from 
improved cooling (squares) and from the spectral flow (diamonds), on a
$12^4$ lattice at $\beta=2.4$. From ref.~[52].\vskip-8mm}
\label{fig:fig6}
\end{figure}

For very smooth instanton fields (instantons of large size), one should
have near continuum behaviour and the crossing of the zero mode should occur
at small values of $m$. Perhaps the crossing value can thus be used to define
the size of the instanton~\cite{Navr}. One way to study the correlation between
size and crossing would be to use cooling to manipulate the size of an 
instanton. Essential is that the spectral flow analysis is done on a coarse 
lattice. No smoothing is necessary. Further studies will be required, but the 
prospects are quite promising. 

\begin{figure}[htb]
\vspace{3.5cm}
\includegraphics{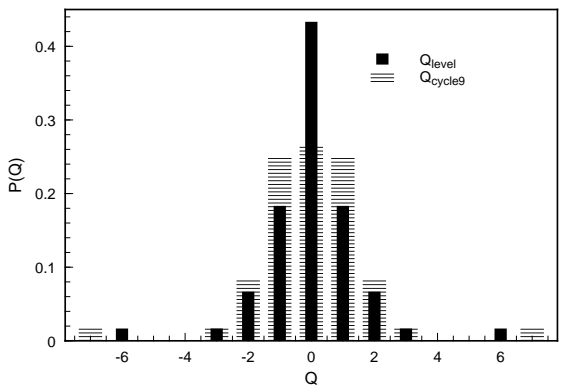}
\caption{Comparison~[52] for the fixed point action of the topological 
charge distribution using the spectral flow ($Q_{\rm level}$) and 
after nine smoothing cycles ($Q_{\rm cycle 9}$) on the same configurations.
\vskip-8mm}
\label{fig:fig7}
\end{figure}

In fig.~6 a comparison is given for the topological charge distribution
generated with the Wilson action on the same lattice, measured with improved 
cooling~\cite{Mgpe} and with the spectral flow~\cite{Navr}. The agreement
is excellent. From the spectral charge one extracts $\chi_t\!=\!(184(6)\MeV)^4$.

Fig.~7 is based on 30 configurations generated with the fixed point action of 
ref.~\cite{Dghk}, on a $12^4$ lattice. The charge measured after nine smoothing
cycles is compared with the spectral charge on the initial rough 
configuration~\cite{Navr}. Cycling is seen to suppress the small charges. 
One configuration was used to trace the cause for this discrepancy~\cite{Navr} 
and it comes as an unpleasant surprise that the spectral charge was 1 without, 
3 after 9 and 2 after 12 smoothing cycles. 

\section{EPILOGUE}

I report - you conclude. Who thought so much can be said about nothing. I 
humbly apologise to those that had hoped to find something else. I would 
have liked to discuss more on finite temperature and implications for the 
instanton liquid, on non-perturbative results in supersymmetric gauge 
theories as a testing ground for QCD and much more. But instantons are 
here to stay.

\section*{Acknowledgements}

I thank the organisers for a job {\em very well} done and for entrusting me 
with the entertainment of the last talk. There are too many I should thank
for discussions and help. Let me make an exception for Margarita Garc\'{\i}a 
P\'erez, Adriano Di Giacomo, Jeff Greensite, Ferenc Niedermayer, Misha Shifman,
Edward Shuryak, Mike Teper and in particular Tom DeGrand and Anna Hasenfratz.

\end{document}